\documentclass[journal]{IEEEtran}

\usepackage{epsfig,amsmath,amssymb,epsf,amsthm,scalefnt,multirow,subfig,lipsum}
\usepackage{xcolor,mathtools,cuted}
\usepackage{float}
\usepackage{cite}
\usepackage{psfrag}
\usepackage{graphics,amsmath,amssymb,graphicx,amsthm,scalefnt,multirow,dsfont,subfig}
\usepackage{algorithm}
\usepackage{algpseudocode}
\usepackage{latexsym,amsmath}
\usepackage{stfloats}
\usepackage{aligned-overset}				

\newcommand{\argmax}{\mathop{\mathrm{argmax}}}
\newcommand{\argmin}{\mathop{\mathrm{argmin}}}

\newtheorem{lemma}{Lemma}
\newtheorem*{lemma*}{Lemma}
\newtheorem*{remark}{Remark}

\def\argmin{\mathop{\mathrm{argmin}}}
\def\argmax{\mathop{\mathrm{argmax}}}

\def\b0{{\pmb{0}}} 

\def\ba{{\mathbf{a}}} \def\bb{{\mathbf{b}}}  
\def\bee{{\mathbf{e}}} \def\bff{{\mathbf{f}}} \def\bg{{\mathbf{g}}} \def\bh{{\mathbf{h}}}
   
   \def\bp{{\mathbf{p}}}
   
  \def\bw{{\mathbf{w}}}

\def\bA{{\mathbf{A}}} \def\bB{{\mathbf{B}}} \def\bC{{\mathbf{C}}} 
 \def\bF{{\mathbf{F}}}  \def\bH{{\mathbf{H}}}
\def\bI{{\mathbf{I}}}   
   
 \def\bR{{\mathbf{R}}}  
  \def\bW{{\mathbf{W}}}

\def\CN{\mathcal{C}\mathcal{N}}

\def\red{\textcolor{black}}

\begin{document}
	
	\title{Complete Power Reallocation for MU-MIMO under Per-Antenna Power Constraint}
	
	\author{Sucheol Kim, Hyeongtaek Lee, Hwanjin Kim, Yongyun Choi, and Junil Choi
		\thanks{S. Kim, H. Lee, H. Kim, and J. Choi are with the School of Electrical Engineering, Korea Advanced Institute of Science and Technology (e-mail: \{loehcumik,htlee8459,jin0903,junil\}@kaist.ac.kr).}
		\thanks{Y. Choi is with the Network Business, Samsung Electronics Co., LTD (e-mail: yongyun.choi@samsung.com).}
	}	
	\maketitle
	
	\begin{abstract}
		This paper proposes a beamforming method under a per-antenna power constraint (PAPC). Although many beamformer designs with the PAPC need to solve complex optimization problems, the proposed complete power reallocation (CPR) method can generate beamformers with excellent performance only with linear operations. CPR is designed to have a simple structure, making it highly flexible and practical. In this paper, three CPR variations considering algorithm convergence speed, sum-rate maximization, and robustness to channel uncertainty are developed. Simulation results verify that CPR and its variations satisfy their design criteria, and, hence, CPR can be readily utilized for various purposes.
	\end{abstract}
	
	\begin{IEEEkeywords}
		Per-antenna power constraint, multi-user multiple-input multiple-output, computational complexity, robust beamforming.
	\end{IEEEkeywords}
	
	\section{Introduction}\label{sec1}
	
	Use of multiple antennas is a popular technique for wireless communication systems \cite{D.Gesbert:2003,A.Goldsmith:2003,A.J.Paulraj:2004}. It allows multi-user (MU) diversity gain to increase \red{the} sum-rate in multi-user multiple-input multiple-output (MU-MIMO) systems \cite{D.Gesbert:2008, A.L.Anderson:2012,A.Wiesel:2006}. Dirty paper coding (DPC) is a well-known transmit technique that achieves the capacity of MU-MIMO downlink channel \cite{G.Caire:2003,H.Weingarten:2004}. DPC, however, requires very high computational complexity to implement in practice, and many linear precoders were developed to lower the computational complexity by feasible level with suboptimal performance instead \cite{T.Yoo:2006,Q.H.Spencer:2004F,B.Li:2015,M.Sadek:2007}. As an attractive linear precoder, a zero-forcing (ZF) beamformer provides fine balance between the complexity and performance~\cite{T.Yoo:2006,Q.H.Spencer:2004F}.
	
	The performance and simplicity of ZF beamformer are guaranteed under the total transmit power constraint but not under the per-antenna power constraint (PAPC). Including the beamformers in \cite{T.Yoo:2006} and \cite{Q.H.Spencer:2004F}, many beamformers were designed under the total transmit power constraint \cite{M.Sadek:2007,M.Joham:2005}. In practice, however, power amplifiers, which have limit on their maximum power, are connected to each antenna \cite{W.Yu:2007,B.Li:2015}. Therefore, the PAPC must be taken into account for practical beamformer designs. 
	
	It is not a simple problem to design beamformers with the PAPC though. Even the ZF beamformer design becomes non-trivial under the PAPC \cite{B.Li:2015}.
	It is possible to simply downscale the ordinary ZF beamformer to satisfy the PAPC; however, this simple approach usually does not exploit full available power at each amplifier, resulting in significant performance degradation \cite{S.Lee:2013}. In \cite{J.Jang:2015}, a beamforming algorithm was proposed to maximize the total transmit power under the PAPC, but no practical communication metrics, e.g., sum-rate, minimum data rate, or beamforming gain, were considered for the beamformer design. \red{By considering a relaxed PAPC, a precoder in \cite{M.Medra:2018-sum_rate_PAPC} is designed to maximize a weighted sum-rate.} In many cases, beamformers were designed by solving one or more optimization problems \red{including} the PAPC as a constraint. 
	In \cite{A.Wiesel:2008}, semidefinite programming (SDP) was adopted to maximize the sum-rate \red{subject to} the PAPC and \red{a} zero interference constraint. The relaxation of rank-one constraint in SDP, however, requires to solve \red{an} additional optimization problem, making the design highly complex.
	The duality of primal- and dual-optimization problems was applied in \cite{W.Yu:2007} and \cite{S.Shen:2014} to maximize the data rate or to minimize the transmit power under the PAPC. In \cite{F.Rashid-Farrokhi:1998}, a normalized beamformer and its power distribution were updated iteratively \red{where one is fixed during the update of another.} Most of the beamformer designs with the PAPC, however, require non-linear operations or solving complicated optimization problems without closed form solutions. The PAPC, which is a must in practice, revives the computational complexity problem that was once relaxed for linear suboptimal beamformers.
	
	Another concern for beamformer designs is whether a base station (BS) has accurate downlink channel information. It would be reasonable that the BS has accurate channel information when the user equipments (UEs) are static, making the wireless channels vary slowly \cite{Q.H.Spencer:2004O,G.J.Foschini:1998}. This is even more feasible for time division duplexing (TDD) systems exploiting uplink/downlink channel reciprocity 
	\cite{J.Hoydis:2013,T.L.Marzentta:2006,L.Lu:2014}.
	We, however, can not expect all the UEs are always static, and uncertainty inheres in the channel information of moving UEs. Therefore, beamformers need to be robust to the channel uncertainty. The beamformer in \cite{Y.Huang:2010} was designed for each UE to minimize the total transmit power ensuring \red{a} quality of service (QoS) with the SDP relaxation under \red{an} interference constraint. 
	In \cite{H.Du:2012}, the signal-to-leakage-plus-noise ratio (SLNR) was maximized with a probabilistic constraint on the leakage power. In \cite{M.B.Shenouda:2007}, the transmit power was minimized under QoS constraints assuming the channel uncertainty as a bounded error. Recently, in \cite{J.Shi:2020}, a deep learning technique was used to generate robust beamformers directly from inaccurate channel information.
	These beamformers do not satisfy the PAPC, making them impractical though. Both the PAPC and robustness to the channel uncertainty were taken into account for the beamformer design in \cite{M.Ding:2013}; however, it requires to solve multiple complicated optimization problems.
	
	In general, linear beamformers lose their benefit of low complexity when the PAPC and the channel uncertainty are taken into account.
	In this paper, we propose a novel low complexity beamformer design, dubbed as complete power reallocation (CPR), which satisfies the PAPC. Proposed CPR iteratively updates a beamformer to fully exploit the transmit power in each antenna while the update can be conducted by linear operations with low complexity.	
	CPR also requires only a finite number of iterations regardless of its design parameters, whereas most of iterative beamformer design algorithms may suffer from a convergence issue depending on their design parameters \cite{M.Ding:2013,F.Rashid-Farrokhi:1998,Y.Huang:2010}.
	With its simple structure, CPR easily can have different variations to aim certain purposes. As concrete examples, we explain three variations of CPR, each for the convergence speed of algorithm, sum-rate maximization, and robustness to the channel uncertainty of moving UEs.
		
	The rest of paper is organized as follows. In Section~\ref{sec2}, system and channel models are described. In Section~\ref{sec3}, the details of proposed CPR and its variations are explained. The numerical results to evaluate the proposed techniques are shown in Section~\ref{sec4}, and the conclusion follows in Section~\ref{sec5}.
	
	$\textbf{Notations}$: 
	A matrix and a vector are written in a bold face capital letter and a bold face small letter. For a matrix or a vector, its transpose, Hermitian, and element-wise conjugate are represented as $(\cdot)^\mathrm{T}$, $(\cdot)^\mathrm{H}$, and $(\cdot)^*$. $\bA^\dagger$~means the pseudo-inverse of matrix~$\bA$.
	The $a$-th column and $b$-th row of matrix $\bA$ are denoted as $(\bA)_{(:,a)}$ and $(\bA)_{(b,:)}$.
	The $b$-th component of vector $\ba$ is remarked as $(\ba)_{(b)}$. $\bI_a$ is the $a\times a$ identity matrix, and $\boldsymbol{0}_a$ represents the $a\times1$ all zero vector.
	The concatenation of matrices is expressed as $[\bA, \bB]$ where $\bA$ and $\bB$ have the same number of rows.
	$\odot$ represents the Hadamard product. 
	A diagonal matrix, of which the diagonal elements are the elements of $\ba$, is denoted as $\mathrm{diag}(\ba)$. \red{$\mathrm{Real}\{a\}$ and $\mathrm{Imag}\{a\}$ are the real and imaginary part of complex number $a$, and the phase of $a$ is represented as $\angle a$}. The uniform distribution in the range $[a,b]$ is written as $\mathrm{U}[a,b]$. The complex multivariate normal distribution with mean vector $\boldsymbol{\mu}$ and covariance matrix $\boldsymbol{\Sigma}$ is represented as $\mathcal{CN}(\boldsymbol{\mu},\boldsymbol{\Sigma})$. For a set $\mathcal{I}$, the function $\mathrm{idx}_{\mathcal{I}}(i)$, $i\in\mathcal{I}$ gives the index of $i$ in the set $\mathcal{I}$, and its cardinality is written as $\mathcal{C}(\mathcal{I})$. A subspace that is spanned by the columns of matrix $\bA$ is denoted as $\mathcal{S}(\bA)$. \red{The function $\mathrm{proj}(\ba,\mathcal{S}(\bA))$ projects a vector $\ba$ on the subspace $\mathcal{S}(\bA)$.}

	\section{System and channel models}\label{sec2}

	We consider an MU-MIMO system with a BS equipped with $M$ antennas and $K$ UEs each with single antenna. Assuming the block-fading model, the received signal at the $k$-th UE during the $i$-th fading block is
	\begin{align}\label{recieved signal}
		y_{k,i} = \bh_{k,i}^\mathrm{H}\bff_{k,i} s_{k,i} + \sum_{\substack{\ell=1\\\ell \neq k}}^K\bh_{k,i}^\mathrm{H}\bff_{\ell,i} s_{\ell,i} + n_{k,i},
	\end{align}
	where $\bh_{k,i}\in\mathbb{C}^{M\times1}$ is the channel vector between the BS and the $k$-th UE, $\bff_{k,i}\in\mathbb{C}^{M\times1}$ is the beamforming vector of BS to support the $k$-th UE, $s_{k,i}$ is the $i$-th transmit symbol\footnote{The transmit symbol $s_{k,i}$ is able to vary during a fading block in practice. \red{To make notation simple, we neglect this fact since the transmit symbol is irrelevant to linear precoder designs.}} for the $k$-th UE satisfying $\mathbb{E}[|s_{k,i}|^2]=1$, and $n_{k,i}\sim \CN(0,\sigma^2)$ is the noise of which variance is the same for all time instances.
	
	For the beamforming vectors $\bff_{k,i}$, we consider practical power constraints as
	\begingroup
	\allowdisplaybreaks
	\begin{align}
		&\sum_{k=1}^{K}\left\lVert \bff_{k,i} \right\rVert_2^2 \leq P_\mathrm{tot}, 	\label{total power constraint}
		\\
		\max_{m\in\{1,\cdots,M\}} &\sum_{k=1}^{K} \left\lvert (\bff_{k,i})_{(m)} \right\rvert^2 \leq P_\mathrm{ant}, 
		\label{PAPC}
	\end{align}
	\red{where \eqref{total power constraint} is the total power constraint with the maximum total transmit power $P_\mathrm{tot}$, and \eqref{PAPC} is the per-antenna power constraint (PAPC) with the maximum antenna transmit power $P_\mathrm{ant}$. For each transmission, the two power constraints need to be satisfied simultaneously. In this paper, we consider the maximum antenna transmit power as $P_\mathrm{ant}={P_\mathrm{tot}}/{M}$, and this let the PAPC in \eqref{PAPC} be a sufficient condition for the total transmit power constraint in \eqref{total power constraint}.}
	\endgroup
		
	The channel vector of each UE is modeled as \cite{H.Kim:2019, H.Q.Ngo:2013}
	\begin{align}\label{channel model}
		\bh_{k,0}&=\bR_k^\frac{1}{2} \bg_{k,0}, 
		\\
		\bh_{k,i}&=\eta_k\bh_{k,i-1}+\sqrt{(1-\eta_k^2)}\bR_k^\frac{1}{2}\bg_{k,i}, \quad i\geq 1,
	\end{align}
	where $\bR_k=\mathbb{E}\left[\bh_{k,i}\bh_{k,i}^\mathrm{H}\right]$ is the spatial correlation matrix, $\bg_{k,i}\sim\CN(\boldsymbol{0}_M,\beta_k\bI_M)$ is the innovation process of the channel vector of the $k$-th UE at the $i$-th fading block, $\beta_k$ models the large-scale fading effect, and $\eta_k$ is the temporal correlation coefficient. At the $i$-th fading-block, the overall channel matrix becomes $\bH_i = [\bh_{1,i}, \cdots, \bh_{K,i}]$ by concatenating the channel vectors of all UEs. 
	For the long-term second order statistics $\bR_k$ and $\eta_k$, we adopt the exponential correlation model and the Jakes' model, respectively \cite{J.Choi:2012}. The exponential spatial correlation matrix is given by 
	\begin{align}\label{spatial correlation}
		\bR_k = 
		\begin{bmatrix}
			1				&r_k			&\cdots	&r_k^{(M-1)}\\
			r_k^*			&1				&\cdots	&r_k^{(M-2)}\\
			\vdots			&\vdots			&\ddots	&\vdots\\
			({r^*_k})^{(M-1)} &(r_k^*)^{(M-2)}	&\cdots	&1
		\end{bmatrix},
	\end{align}
	where $r_k\in\mathbb{C}$ satisfies $|r_k| < 1$ and $0\leq \angle r_k <2\pi$. The Jakes' model for the temporal correlation is given as 
	\begin{align}
		\eta_k=J_0(2\pi f_{D,k}t),
	\end{align}
	where $J_0(\cdot)$ is the $0$-th order Bessel function, $f_{D,k}$ is the Doppler frequency, and $t$ is the channel instantiation interval. The Doppler frequency of a static UE is zero, and the channel vector is invariant over $i$ with $\eta_k=1$. For a moving UE, the temporal correlation becomes less than one, and the corresponding channel vector varies over time.
	
	\section{Beamformer designs}\label{sec3}
	
	The motivation and concept of the proposed CPR method are introduced in Section \ref{sec3-a}. Then, the details of CPR and its variations are explained from Section \ref{sec3-b} to Section \ref{sec3-e} aiming for different purposes or circumstances. To explain CPR clearly, we assume all UEs are static from Section \ref{sec3-a} to Section \ref{sec3-d}, and the block-fading index $i$ of channels and beamformers is omitted for legibility. In Section \ref{sec3-e}, we consider several moving UEs and restore the time index $i$ to distinguish the outdated and the current channels.
	
	\subsection{Motivation and concept of CPR}\label{sec3-a}
		
	For a single UE multiple-input single-output (MISO) system, a beamformer that maximizes the beamforming gain is the matched beamformer. Under the power constraints in \eqref{total power constraint} and \eqref{PAPC}, the matched beamformer can be designed as
	\begin{align}\label{scaled matched beamformer}
		\bff_\mathrm{sMB} &= \sqrt{P_\mathrm{ant}}\frac{\bh}{\left\lvert(\bh)_{(m_{\max})} \right\rvert},
		\\
		m_\mathrm{max} &= \argmax_{m\in\{1,\cdots,M\}} \left\lvert (\bh)_{(m)} \right\rvert,
	\end{align}
	where $\bh\in\mathbb{C}^{M\times1}$ is the MISO channel. When only the total transmit power constraint is considered, the ordinary matched beamformer $\bff_\mathrm{MB}=\sqrt{P_\mathrm{tot}}\frac{\bh}{\left\lVert\bh\right\rVert_2}$ maximizes the beamforming gain $|\bh^\mathrm{H}\bff_\mathrm{MB}|^2$. However, the beamformer in \eqref{scaled matched beamformer}, which is additionally constrained by the PAPC, is a scaled-down version of the ordinary matched beamformer, and the transmit power of the beamformer $\bff_\mathrm{sMB}$ is
	\begin{align}
		\left\lVert\bff_\mathrm{sMB}\right\rVert_2^2 
		= 
		P_\mathrm{ant}\frac{\left\lVert\bh\right\rVert_2^2}{\left\lvert(\bh)_{(m_{\max})}\right\rvert^2}
		\le
		\left\lVert\bff_\mathrm{MB}\right\rVert_2^2
		=
		P_\mathrm{tot},
	\end{align}
	where the equality scarcely holds when all the elements of $\bff_\mathrm{MB}$ have the same magnitude. The decrease of transmit power of beamformer $\bff_\mathrm{sMB}$ depends on the distribution of magnitude of elements $|(\bff_\mathrm{MB})_{(m)}|$. With widespread of magnitude $|(\bff_\mathrm{MB})_{(m)}|$, some \red{antennas} may need to suppress their transmit power significantly, resulting in \red{a} low beamforming gain.
	
	The scaled-down beamformer $\bff_\mathrm{sMB}$ usually has far less transmit power than the total transmit power, i.e., $\left\lVert\bff_\mathrm{sMB}\right\rVert_2^2 < P_\mathrm{tot}$, by the reason that the power of most elements is below the maximum antenna power $|(\bff_\mathrm{sMB})_{(m)}|^2<P_\mathrm{ant}$. To increase the transmit power of scaled beamformer, we can take advantage of the gap between the maximum antenna power $P_\mathrm{ant}$ and the power of the beamformer elements $|(\bff_\mathrm{sMB})_{(m)}|^2$. If we can design a beamformer $\bff'$ that fully exploits the gap, i.e., $|(\bff')_{(m)}|^2=P_\mathrm{ant},\ \forall m\in\{1,\cdots,M\}$ and $\left\lVert\bff'\right\rVert_2^2=P_\mathrm{tot}$, the beamforming gain may largely increase. 
	
	As one possible approach, we can add an extra beamformer $\bw\in\mathbb{C}^{M\times1}$ to the scaled beamformer
	\begin{align}\label{Eq. sMB addition}
		\bff' &= \bff_\mathrm{sMB} + \bw.
	\end{align}
	The design of $\bw$, then, is to maintain the property of the matched beamformer and to make the best use of the antenna power as
	\begin{align}
		(\bw)_{(m)} = \left(\sqrt{P_\mathrm{ant}} - |(\bff_\mathrm{sMB})_{(m)}|\right) e^{j\angle(\bff_\mathrm{sMB})_{(m)}},
	\label{Eq. extra for MB}
	\end{align}
	where $m\in\{1,\cdots,M\}$.	The resulting beamformer $\bff'$ uses the maximum total transmit power, and each antenna element also uses the maximum antenna power. In fact, the beamformer $\bff'$ is the optimal solution of the following optimization problem
	\begin{align}
		\underset{\bff\in\mathbb{C}^{M\times1}}{\text{maximize }} & \left\lvert \bh^\mathrm{H} \bff \right\rvert^2
		\notag
		\\
		\text{subject to }& \left\lVert \bff \right\rVert_2^2 \leq P_\mathrm{tot}
		\notag
		\\
		&\left\lvert (\bff)_{(m)} \right\rvert^2 \leq P_\mathrm{ant},\quad {m\in\{1,\cdots,M\}},
		\label{Eq. MISO optimization with PAPC}
	\end{align}
	which finds the beamformer that maximizes the beamforming gain under the transmit power constraints in \eqref{total power constraint} and \eqref{PAPC}. \red{This can be shown by expanding the objective function in \eqref{Eq. MISO optimization with PAPC} as
	\begin{align}
		\left\lvert \bh^\mathrm{H}\bff \right\rvert^2 
		&= \left\lvert \sum_{m=1}^M (\bh)_{(m)}^*(\bff)_{(m)} \right\rvert^2
		\notag \\
		&= \left\lvert \sum_{m=1}^M \left\lvert(\bh)_{(m)} \right\rvert e^{-j\angle (\bh)_{(m)}} \left\lvert (\bff)_{(m)} \right\rvert e^{j\angle(\bff)_{(m)}} \right\rvert^2
		\notag\\
		\overset{(a)}&{\le} \left\lvert \sum_{m=1}^M \left\lvert(\bh)_{(m)} \right\rvert  \left\lvert (\bff)_{(m)} \right\rvert  \right\rvert^2
		\notag\\
		\overset{(b)}&{\le} P_\text{ant} \left\lvert \sum_{m=1}^M \left\lvert(\bh)_{(m)} \right\rvert \right\rvert^2,
		\label{Eq. MISO optimum}
	\end{align}
	where the equalities in $(a)$ and $(b)$ can be achieved by setting $\angle(\bff)_{(m)}=\angle(\bh)_{(m)}$ and $\left\lvert(\bff)_{(m)}\right\rvert=\sqrt{P_\text{ant}}$, respectively, which are the conditions of optimal solution. The extra beamformer $\bw$ in \eqref{Eq. extra for MB} let the combined beamformer $\bff'$ in \eqref{Eq. sMB addition} satisfy the two equality conditions in \eqref{Eq. MISO optimum} and makes $\bff'$ optimal for \eqref{Eq. MISO optimization with PAPC}.}

	\red{For a MISO system, both the beamformer in \eqref{Eq. sMB addition} and the result of optimization problem in \eqref{Eq. MISO optimization with PAPC} provide the same optimal beamformer. Fortunately, the problem \eqref{Eq. MISO optimization with PAPC} has a closed form solution, which makes it quite easy to build the optimal beamformer satisfying the PAPC. For an MU-MIMO system, however, the PAPC with multiple beamformers is hard to be analyzed, and this makes it difficult to obtain a good beamformer satisfying the PAPC. The focus of this paper is to develop CPR, a low complexity MU-MIMO beamformer design method, under the same motivation and concept of the beamformer design in \eqref{Eq. sMB addition}. 
	}
	
	\subsection{Complete power reallocation (CPR)}\label{sec3-b}

	We now consider the MU-MIMO system in which all UEs are static. To support multiple UEs, it is possible to adopt the same approach in Section \ref{sec3-a} to the beamformer for each UE. The power constraints in \eqref{total power constraint} and \eqref{PAPC}, however, jointly affect beamformers for all UEs, and the power increase of beamformer of a specific UE may reduce the available power of beamformers for other UEs. In addition, the matched beamformer discussed in Section \ref{sec3-a} may result in larger interference for other UEs with increased transmit power. We, hence, need an effective beamformer to handle the inter-user interference and to distribute the transmit power over UEs while satisfying the power constraints in \eqref{total power constraint} and \eqref{PAPC}.

	Proposed CPR develops a beamformer by combining multiple effective beamformers that can manage interference, e.g., ZF beamformer, SLNR beamformer \cite{M.Sadek:2007}, or regularized zero-forcing (RZF) beamformer \cite{C.B.Peel:2005}. In this paper, the ZF beamformer is adopted as a concrete example to explain CPR. 

	The outline of ZF-based CPR is shown in Algorithm \ref{CPR algorithm}. For the given initial beamformer $\bF_0$ and algorithm parameter $p$, CPR iteratively adds up extra beamformers. At each iteration, the extra beamformer $\widehat{\bW}_n$ is obtained for the antenna subset $\mathcal{I}_n^{(p)}$, of which elements have power less than $pP_\mathrm{ant}$. After the dimension of the extra beamformer $\widehat{\bW}_n$ is properly set to form $\bW_n$, it is added to the previous beamformer $\bF_{n-1}$ after adjusting the power distribution with the coefficient matrix $\bA_n=\mathrm{diag}(\ba_n)$. In this subsection, $\bA_n$ is designed for the equal power distribution over UEs while another power distribution strategy for unequal power distribution will be discussed in Section \ref{sec3-d}. For the equal power distribution, we set $\ba_n$ as
	\begin{align}
		(\ba_n)_{(k)} = \alpha_{(\mathrm{EP},n)} \frac{1}{\left\lVert (\bW_n)_{(:,k)} \right\rVert_2}, \quad k\in\{1,\cdots,K\},
		\label{a vector}
	\end{align}
	where $\alpha_{(\mathrm{EP},n)}$ is the coefficient to satisfy the PAPC.

	\red{Under the PAPC, $\alpha_{(\mathrm{EP},n)}$ is designed to maximize the sum-rate of the updated beamformer as
	\begin{align}\label{alpha optimization}
		\underset{\hat{\alpha}_n\in\mathbb{C}}{\text{argmax }} &\sum_{k=1}^K \log_2\left(1+\frac{1}{\sigma}\left\lvert \bh_k^\mathrm{H}\left(\bF_{n-1}+\bW_n\bA_n\right)_{(:,k)}\right\lvert^2\right)
		\notag
		\\
		\text{subject to } & \sum_{k=1}^{K}\left\lVert (\bF_{n-1}+ \bW_n \bA_n)_{(:,k)} \right\rVert_2^2 \leq P_\mathrm{tot}, 
		\notag
		\\
		&\max_{m\in\{1,\cdots,M\}} \sum_{k=1}^{K} \left\lvert (\bF_{n-1}+\bW_n \bA_n)_{(m,k)} \right\rvert^2 \leq P_\mathrm{ant}, 
		\notag
		\\
		&\angle\left(\bh_k^\mathrm{H}(\bF_{n-1})_{(:,k)}\right)=\angle\left(\bh_k^\mathrm{H}(\bW_n\bA_n)_{(:,k)}\right),
		\notag\\
		&\qquad\qquad\qquad\qquad\qquad\qquad
		\quad 
		k\in\{1,\cdots,K\},
		\notag
		\\
		& \bA_n = \mathrm{diag}(\ba_n),
		\notag
		\\
		& (\ba_n)_{(k)} = \hat{\alpha}_n \frac{1}{\left\lVert (\bW_n)_{(:,k)} \right\rVert_2}, \quad k\in\{1,\cdots,K\},
	\end{align}
	where the third constraint aligns the product of the channel vector with the previous beamformer $\bh_k^\mathrm{H}(\bF_{n-1})_{(:,k)}$ and that with the $n$-th extra beamformer $\bh_k^\mathrm{H}(\bW_n\bA_n)_{(:,k)}$. The two beamformers $(\bF_{n-1})_{(:,k)}$ and $(\bW_n)_{(:,k)}$ provide positive real values when they are multiplied with channel vector $\bh_k$, and this restricts the coefficient $\hat{\alpha}_n$ in \eqref{alpha optimization} also to be a real value. Then, the data rate maximization of the $k$-th UE becomes the same as the maximization of the coefficient $\hat{\alpha}_n$ as
	\begin{align}
		&\argmax_{\hat{\alpha}_n\in\mathbb{R}} \log_2\left(1+\frac{1}{\sigma}\left\lvert \bh_k^\mathrm{H}\left(\bF_{n-1}+\bW_n\bA_n\right)_{(:,k)}\right\lvert^2\right)
		\notag\\
		&=\argmax_{\hat{\alpha}_n\in\mathbb{R}} \left\lvert \bh_k^\mathrm{H}\left(\bF_{n-1}\right)_{(:,k)}\right\rvert^2+\left\lvert\bh_k^\mathrm{H}\left(\bW_n\bA_n\right)_{(:,k)}\right\lvert^2
		\notag\\
		&\quad +2\cdot\mathrm{Real}\left\{\bh_k^\mathrm{H}\left(\bF_{n-1}\right)_{(:,k)}\left(\bh_k^\mathrm{H}\left(\bW_n\bA_n\right)_{(:,k)}\right)^*\right\}
		\notag\\
		\overset{(a)}&{=}\argmax_{\hat{\alpha}_n\in\mathbb{R}} \frac{\hat{\alpha}_n^2}{\lVert\left(\bW_n\right)_{(:,k)}\rVert_2^2}\left\lvert\bh_k^\mathrm{H}\left(\bW_n\right)_{(:,k)}\right\lvert^2
		\notag\\
		&\quad
		+\frac{2\hat{\alpha}_n}{\lVert\left(\bW_n\right)_{(:,k)}\rVert_2^2}\underbrace{\bh_k^\mathrm{H}\left(\bF_{n-1}\right)_{(:,k)}\left(\bh_k^\mathrm{H}\left(\bW_n\right)_{(:,k)}\right)^*}_\text{positive real number}
		\notag\\
		&=\argmax_{\hat{\alpha}_n\in\mathbb{R}} \hat{\alpha}_n, \label{Eq. modified alpha objective function}
	\end{align}
	where $(a)$ is by the alignment constraint in \eqref{alpha optimization}. Since the objective function in \eqref{Eq. modified alpha objective function} is independent of the UE index $k$, we can design the single variable $\hat{\alpha}_n$ to maximize the data rate of each UE, which directly maximizes the sum-rate.}

\begin{algorithm}[!t]
	\caption{CPR algorithm}
	\label{CPR algorithm}
	\textbf{Initialize}
	\begin{algorithmic}[1]
		\State Set initial beamformer: 
		\begin{align*}
			\bF_0 = \left[\bff_1, \cdots, \bff_K \right] = \left[\boldsymbol{0}_{M}, \cdots, \boldsymbol{0}_{M} \right]
		\end{align*}
		\State Set parameter $0< p\le 1$ and stopping criteria
	\end{algorithmic}
	\textbf{Beamformer update}
	\begin{algorithmic}[1]
		\addtocounter{ALG@line}{+2}		
		\State \textbf{For} $1\le n \le \red{M-K+1}$
		\State Find antenna set: 
		\begin{align*}
			\mathcal{I}_n^{(p)} = \left\{ m: \sum_{k=1}^{K}\left\lvert (\bff_k)_{(m)} \right\rvert^2 < p P_\mathrm{ant} \right\}
		\end{align*}
		\State \red{\textbf{If} $\mathcal{C}(\mathcal{I}_n^{(p)})<M$:}
		\begin{align*}
			\red{\textbf{break}}
		\end{align*}
		\State \red{\textbf{End if}}
		\State Calculate an extra beamformer for the antenna set: 
		\begin{align*} 
			\widehat{\bW}_n &= \left(\widehat{\bH}_n^\mathrm{H}\right)^\dagger \in \mathbb{C}^{\mathcal{C}(\mathcal{I}_n^{(p)})\times K}
			\notag\\
			\widehat{\bH}_n&=(\bH)_{(\mathcal{I}_n^{(p)},:)}
		\end{align*}
		\State Adjust beamformer dimension:
		\begin{align*}
			(\bW_n)_{(m,:)} = 
			\begin{cases}
				(\widehat{\bW}_n)_{(\mathrm{idx}_{\mathcal{I}_{n}^{(p)}}(m),:)}, & m\in\mathcal{I}_n^{(p)}
				\\
				\boldsymbol{0}_K^\mathrm{T}, & m\notin\mathcal{I}_n^{(p)}
			\end{cases}
		\end{align*}
		\State Set coefficient matrix:
		\begin{align*}
			\bA_n = \mathrm{diag}(\ba_n)
		\end{align*}
		\State Combine beamformers:
		\begin{align*}
			\bF_n = \bF_{n-1} + \bW_n \bA_n
		\end{align*}
		\State \textbf{If} one of stopping criteria is satisfied: 
		\begin{align*}
			&\bF_{\mathrm{CPR}} = \bF_n \\
			&\textbf{break}
		\end{align*}
		\State \textbf{End if}
		\State \textbf{End for}
	\end{algorithmic}
\end{algorithm}
\begin{figure*}[t]
	\centering
	\begin{align}
		&\left\lvert\hat{\alpha}_{n,m}\right\rvert^2 \left\lVert (\bW_n)_{(m,:)}\odot\overline{\bw}_n\right\rVert_2^2 
		+ 2\cdot\mathrm{Real}\left\{\hat{\alpha}_{n,m}  (\bF_{n-1})_{(m,:)}^*\left((\bW_n)_{(m,:)}\odot\overline{\bw}_n\right)^\mathrm{T}\right\} + \left\lVert (\bF_{n-1})_{(m,:)}\right\rVert_2^2 - P_\mathrm{ant} = 0 
		\tag{19}\label{alpha second order equation}
		\\
		&\hat{\alpha}_{n,m}= 
		\frac{1}{\left\lVert\left(\bW_n\right)_{(m,:)}\odot\overline{\bw}_n\right\rVert_2^2}
		\Bigg( -\left(\bF_{n-1}\right)_{(m,:)}\left(\left(\bW_n\right)_{(m,:)}\odot\overline{\bw}_n\right)^\mathrm{H}
		\notag\\
		&
		\qquad\quad
		+e^{j\theta_m}\sqrt{ \left\lvert\left(\bF_{n-1}\right)_{(m,:)}\left(\left(\bW_n\right)_{(m,:)}\odot\overline{\bw}_n\right)^\mathrm{H}\right\rvert^2
			-\left\lVert\left(\bW_n\right)_{(m,:)}\odot\overline{\bw}_n\right\rVert_2^2 \left( \left\lVert \left( \bF_{n-1}\right)_{(m,:)}\right\rVert_2^2-P_\mathrm{ant}\right) }			
		\Bigg)
		\tag{20}\label{alpha closed form}
		\\
		&\red{\theta_m = \sin^{-1}\left(\frac{\mathrm{Imag}\left\{\left(\bF_{n-1}\right)_{(m,:)}\left(\left(\bW_n\right)_{(m,:)}\odot\overline{\bw}_n\right)^\mathrm{H}\right\}}{
				\sqrt{ \left\lvert\left(\bF_{n-1}\right)_{(m,:)}\left(\left(\bW_n\right)_{(m,:)}\odot\overline{\bw}_n\right)^\mathrm{H}\right\rvert^2
					-\left\lVert\left(\bW_n\right)_{(m,:)}\odot\overline{\bw}_n\right\rVert_2^2 \left( \left\lVert \left( \bF_{n-1}\right)_{(m,:)}\right\rVert_2^2-P_\mathrm{ant}\right)}			
			}\right)
			\tag{21}\label{Eq. theta_m}}
	\end{align}
\hrule
\end{figure*}

\begingroup\allowdisplaybreaks
	\red{By replacing the objective function in \eqref{alpha optimization} with the magnitude of coefficient $|\hat{\alpha}_n|$, we can obtain the optimal solution in a closed form. To find the solution $\alpha_{(\mathrm{EP},n)}$, we first find the possible values $\hat{\alpha}_{n,m}$ that satisfy the PAPC of each $m$-th antenna with equality
	\begin{align}
		\sum_{k=1}^{K} \left\lvert (\bF_{n-1})_{(m,k)}+\frac{\hat{\alpha}_{n,m}}{\left\lVert(\bW_{n})_{(:,k)}\right\rVert_2}(\bW_n)_{(m,k)} \right\rvert^2 &= P_\mathrm{ant},
	\end{align}
	which can be reformulated as a second order complex equation as \eqref{alpha second order equation} 
	on the top of the next page, where
	$\overline{\bw}_n=\left[\left\lVert(\bW_n)_{(:,1)}\rVert_2^{-1},\cdots,\right\lVert(\bW_n)_{(:,K)}\rVert_2^{-1}\right]$. The root of the second order equation is in \eqref{alpha closed form}, and we set $\theta_m$ to align 
	the two beamformers $\bF_{n-1}$ and $\bW_n$ as in \eqref{Eq. theta_m}.}
	Then, the solution of the problem \eqref{alpha optimization}, which needs to satisfy the PAPC for all antennas, is obtained as
	\setcounter{equation}{21}
	\begin{align}\label{alpha final}
		\alpha_{(\mathrm{EP},n)} = \argmin_{\hat{\alpha}_n\in\{\hat{\alpha}_{n,1},\cdots,\hat{\alpha}_{n,m}\}}
		|\hat{\alpha}_n|.
	\end{align}
	\red{With the maximum antenna power $P_\mathrm{ant}=P_\mathrm{tot}/M$, the updated beamformer satisfies the total transmit power constraint when it satisfies the PAPC as
	\begin{align}
		&
		\sum_{k=1}^{K}\left\lVert (\bF_{n-1}+ \bW_n \bA_n)_{(:,k)} \right\rVert_2^2 
		\notag\\
		&= \sum_{m=1}^M \sum_{k=1}^K\left\lvert (\bF_{n-1}+ \bW_n \bA_n)_{(m,k)} \right\rvert^2
		\notag\\
		&\leq \sum_{m=1}^M P_\mathrm{ant}
		\notag\\
		&= P_\mathrm{tot}.
	\end{align}}
\endgroup

\begin{figure}[t]
		\centering
		\includegraphics[width=1\linewidth]{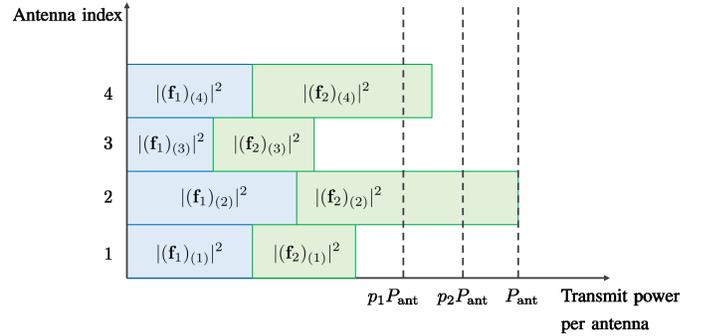}
		\caption{Transmit power per-antenna with $M=4$, $K=2$.}
		\label{TxP per ant}
	\end{figure}

	\red{The design of coefficient matrix $\bA_n$ in \eqref{alpha optimization} makes at least one antenna in the antenna set $\mathcal{I}_n^{(p)}$ to use the maximum antenna power $P_\mathrm{ant}$ at each algorithm iteration. As the proposed algorithm iterates, then, the number of antennas that use the maximum antenna power increases. Since Line~5 in Algorithm~\ref{CPR algorithm} requires the cardinality of antenna set to be no smaller than the number of UEs $\mathcal{C}(\mathcal{I}_n^{(p)})\ge K$, the maximum algorithm iteration is restricted by $M-K+1$. Under the maximum algorithm iteration, the parameter $p$ in Algorithm~\ref{CPR algorithm}} affects the algorithm convergence speed and the transmit power of final beamformer $\bF_{\mathrm{CPR}}$. A small value of $p$ would increase the magnitude of $(\ba_n)_{(k)}$, speeding up the convergence of CPR algorithm. As a simple example, Fig. \ref{TxP per ant} represents the per-antenna transmit power of a beamformer $\bF=[\bff_1, \bff_2]$ with $M=4$ BS antennas and $K=2$ UEs. The second antenna is using the maximum antenna power $P_\mathrm{ant}$, and if we take a small $p=p_1$ and $\bF_{n-1}=\bF$, the antenna set for iteration in Line 4 of Algorithm \ref{CPR algorithm} becomes $\mathcal{I}_n^{(p_1)} = \{1,3\}$. The corresponding extra beamformer after dimension adjustment is
	\begin{align}
		\bW_n = 
		\begin{bmatrix}
			(\widehat{\bW}_n)_{(1,:)}
			\\
			\boldsymbol{0}_K^\mathrm{T}
			\\
			(\widehat{\bW}_n)_{(2,:)}
			\\
			\boldsymbol{0}_K^\mathrm{T}
		\end{bmatrix},
	\end{align}
	where $\widehat{\bW}_n\in\mathbb{C}^{\mathcal{C}(\mathcal{I}_n^{(p_1)})\times K}$ is the extra beamformer before the adjustment. 
	The expected magnitude of $(\ba_n)_{(k)}$ is approximately proportional to the gap $P_\mathrm{ant} - \max_{m\in\mathcal{I}_n^{(p_1)}}\{(|(\bff_1)_{(m)}|^2+|(\bff_2)_{(m)}|^2\}$, which represents the amount of available power not in use. On the contrary, if we take relatively large $p=p_2$ and $\bF_{n-1}=\bF$, the antenna set becomes $\mathcal{I}_n^{(p_2)}=\{1,3,4\}$, and the magnitude of $(\ba_n)_{(k)}$ is approximately proportional to a smaller gap $P_\mathrm{ant} - \max_{m\in\mathcal{I}_n^{(p_2)}}\{(|(\bff_1)_{(m)}|^2+|(\bff_2)_{(m)}|^2\}$ than the case of $p=p_1$. With small $p=p_1$, large magnitude of coefficients $|(\ba_n)_{(k)}|$ raises the transmit power increment at each iteration and consequently accelerates the convergence speed of CPR algorithm by rapidly increasing the transmit power \red{close} to $P_\mathrm{tot}$. With small value $p=p_1$, however, it is not possible to exploit the potential power of 4-th antenna $P_\mathrm{ant} - (|(\bff_1)_{(4)}|^2+|(\bff_2)_{(4)}|^2)$. Hence, we can balance the total transmit power of the final beamformer $\bF_{\mathrm{CPR}}$ and the convergence speed of CPR algorithm by selecting \red{a} proper $p$.
	
%

	While the maximum iteration number of CPR is \red{${M-K+1}$} as stated in Line 3 of Algorithm~\ref{CPR algorithm}, we can set specific stopping criteria considering the power constraints in \eqref{total power constraint} and \eqref{PAPC}. For example, Algorithm~\ref{CPR algorithm} can be stopped when 99\% of total transmit power is used or 90\% of antennas are exploiting the full antenna power $P_\mathrm{ant}$. \red{Depending on the beamformers used in CPR, we may need to consider other stopping criteria. For CPR with the ZF beamformer, in Algorithm~\ref{CPR algorithm}, we included Line 5 to ensure the full column rank of matrix $\widehat{\bH}_n$ in Line 7, which can be satisfied when $\mathcal{C}(\mathcal{I}_n^{(p)})\ge K$ as long as all UEs experience independent channels.} It is obvious this criterion can be met faster with small value $p=p_1$ than with large value $p=p_2$.
	
	\red{While CPR can accommodate any linear beamformers by adjusting Line 7, the use of ZF beamformer provides a benefit that the final beamformer $\bF_{\mathrm{CPR}}$ has a higher beamforming gain than the initial beamformer without additional interference. This is shown in the following lemma.
	\begin{lemma}\label{ZF lemma}
		For an antenna set $\mathcal{I}_n^{(p)}$, a beamformer $\bF_{n-1}\in\mathbb{C}^{M\times K}$, and two positive integers $M$ and $K$, if an extra beamformer $\widehat{\bW}_n\in\mathbb{C}^{\mathcal{C}(\mathcal{I}_n^{(p)})\times K}$ is a ZF beamformer of channel $(\bH)_{(\mathcal{I}_n^{(p)},:)}$, then the beamformer
		\begin{align}
			\bF_n = \bF_{n-1} + \bW_n \bA_n, 
		\end{align}
		 gives a strictly increased beamforming gain without additional interference for all UEs where $\bA_n$ and $\bW_n$ are defined in Algorithm~\ref{CPR algorithm}.
	\end{lemma}}
	\begingroup
	\allowdisplaybreaks
	\begin{IEEEproof} 
		\red{See Appendix I}
	\end{IEEEproof}
	\endgroup
	\red{
	Lemma \ref{ZF lemma} holds for any initial beamformers as long as $\widehat{\bW}_n$ is set to be the ZF beamformer. This means that any well-designed beamformer can be improved further by Algorithm~\ref{CPR algorithm}, 
	unless the beamformer satisfies the stopping criteria of Algorithm~\ref{CPR algorithm}.
	In Algorithm~\ref{CPR algorithm}, the initial beamformer is set as a zero matrix, and the following remark states that this initialization let the final beamformer $\bF_{\mathrm{CPR}}$ be a ZF beamformer.
	\begin{remark}
		When the initial beamformer $\bF_0$ is set to be a zero matrix $[\boldsymbol{0}_M,\cdots,\boldsymbol{0}_M]$, the updated beamformers $\bF_n$ in Algorithm~\ref{CPR algorithm} are ZF beamformers with increasing beamforming gains as the algorithm iterates.
	\end{remark}
	Under the maximum iteration number $M-K+1$, Lemma~\ref{ZF lemma} assures the convergence of Algorithm~\ref{CPR algorithm}.
	}

	
	\subsection{Fast convergence CPR (FC-CPR)}\label{sec3-c}

	In practice, it would be better to increase the transmit power as much as possible at early iterations to speed up CPR process. In this subsection, we propose fast convergence CPR (FC-CPR), which is a variation of CPR to have less number of CPR iterations. FC-CPR exploits CPR based on the fact that 1) CPR converges faster with small $p$ in Algorithm~\ref{CPR algorithm} and 2) the increase of transmit power in each CPR process is approximately proportional to the gap between $P_\mathrm{ant}$ and the maximum of current transmit powers allocated to the antennas to be updated as explained in the previous subsection. As in Algorithm~\ref{FC-CPR algorithm}, FC-CPR sets the initial parameter \red{as a small value $p=p_\mathrm{init}$} and increases $p$ by $\Delta_p$ whenever CPR converges. A small $p$ makes the size of the antenna set~$\mathcal{I}_n^{(p)}$ in Line 4 of Algorithm \ref{CPR algorithm} small and allocates much power only to the antennas in $\mathcal{I}_n^{(p)}$, letting CPR converges faster. With small $p$, however, the antennas not included in the set~$\mathcal{I}_n^{(p)}$ may not fully exploit their available power. \text{FC-CPR}, therefore, increases $p$ whenever CPR converges, and then CPR operates with another small set of antennas. This let FC-CPR deal with most of antennas exploiting their full power. FC-CPR also allows to handle a small dimensional channel matrix $(\bH)_{(\mathcal{I}_n^{(p)},:)}$ in Line 7 of Algorithm \ref{CPR algorithm} making overall process less complex. FC-CPR stops when $p\ge p_{\max}$ with a predefined $p_{\max}$.
		
		\begin{algorithm}[!t]
			\caption{FC-CPR algorithm}
			\label{FC-CPR algorithm}
			\textbf{Initialize}
			\begin{algorithmic}[1]
				\State Set initial beamformer: 
				\begin{align*}
					\bF_0 = \left[\bff_1, \cdots, \bff_K \right] = \left[\boldsymbol{0}_{M}, \cdots, \boldsymbol{0}_{M} \right]
				\end{align*}
				\State Set parameters $0 < \red{p=p_\mathrm{init}} < p_{\max}$, $0 < p_{\max} \le 1$, and $\Delta_p>0$
			\end{algorithmic}
			\textbf{Beamformer update}
			\begin{algorithmic}[1]
				\addtocounter{ALG@line}{+2}		
				\State \textbf{For} $1\le n' \le \red{M-K+1}$
				\State Run CPR with initial beamformer $\bF_0$ and $p$
				\State \textbf{If} $p < p_{\max}$ 
				\begin{align*}
					&p = p + \Delta_p\\
					&\bF_0 = \bF_\mathrm{CPR}
				\end{align*}
				\State \textbf{Else}
				\begin{align*}
					& \bF_{\mathrm{FCCPR}} = \bF_{\mathrm{CPR}}\\
					&\textbf{break}
				\end{align*}
				\State \textbf{End if}
				\State \textbf{End for}
			\end{algorithmic}
		\end{algorithm} 
		
	\red{Although FC-CPR exploits CPR multiple times, the total iteration number of FC-CPR is \red{usually} smaller than that of CPR. Note that $\alpha_{(\mathrm{EP},n)}$ in \eqref{a vector} guarantees that at least one antenna in the set $\mathcal{I}_n^{(p)}$ is allocated with the maximum antenna power $P_\mathrm{ant}$ per iteration, and the stopping criteria in Line~5 of Algorithm~\ref{CPR algorithm} also holds for FC-CPR. Therefore, the total iteration number of FC-CPR is always smaller than $M-K+1$, and the large amount of transmit power increment per iteration let FC-CPR converge faster than CPR. This is numerically shown in Section IV.}

	\subsection{Power distribution strategy of CPR}\label{sec3-d}
	
	We have considered $\ba_n$ in \eqref{a vector} to achieve equal power distribution over UEs. It is possible to have non-uniform power distribution strategies by setting $\ba_n$ as
	\begin{align}\label{a vector unequal}
		(\ba_n)_{(k)}=\alpha_{n}\frac{(\bb)_{(k)}}{\left\lVert(\bW_n)_{(:,k)}\right\rVert_2}, \quad k\in\{1,\cdots,K\},
	\end{align}
	where $\bb\in\mathbb{C}^{K\times1}$ determines the power distribution across UEs. Since $\bb$ is fixed over all CPR iterations, $\ba_n$ is still updated by a single variable $\alpha_{n}$ where $\alpha_n$ can be obtained by solving \eqref{alpha optimization} after substituting $(\ba_n)_{(k)} = \hat{\alpha}_n \frac{1}{\left\lVert (\bW_n)_{(:,k)} \right\rVert_2}$ with $(\ba_n)_{(k)} = \hat{\alpha}_n \frac{(\bb)_{(k)}}{\left\lVert (\bW_n)_{(:,k)} \right\rVert_2}$.

	We adopt the water-filling power distribution strategy, which is known as \red{an optimal solution} to maximize the sum-rate, as a representative example. The conventional water-filling problem, however, does not consider the PAPC, and we propose a strategy that mimics the water-filling solution. First, we set $\bb_\mathrm{WF}$ as the water-filling solution of following problem
	\begingroup
	\allowdisplaybreaks 
	\begin{align}\label{b optimization}
		\underset{\hat{\bb}\in\mathbb{C}^{K\times1}}{\text{maximize }} &\sum_{k=1}^K \log_2\left(1+ \frac{|(\hat{\bb})_{(k)}\bh_k^\mathrm{H}\bp_k|^2}{\sigma^2+\left\lvert \sum_{\ell\neq k} (\hat{\bb})_{(\ell)}\bh_k^\mathrm{H}\bp_\ell \right\rvert^2} \right) 
		\notag
		\\
		\text{subject to } & \bp_k =  \frac{\left(\left(\bH^\mathrm{H}\right)^\dagger\right)_{(:,k)}}{\left\lVert\left(\left(\bH^\mathrm{H}\right)^\dagger\right)_{(:,k)}\right\rVert_2}, \quad k\in\{1,\cdots,K\},
		\notag
		\\
		& \sum_{k=1}^{K}|(\hat{\bb})_{(k)}|^2 \leq P_\mathrm{tot},
	\end{align}
	\endgroup
	where the problem has a well-known closed form solution \red{\cite{J.Jang:2003_WF_solution}. We let the optimal solution of \eqref{b optimization} as $\bb_{\mathrm{WF}}$ that is given as
	\begin{align}
		(\bb_{\mathrm{WF}})_{(k)}=\sigma\max\left\{0,\sqrt{\frac{1}{\lambda}-\frac{1}{\left\lvert\bh_k^\mathrm{H}\bp_k\right\rvert^2}}\right\},
	\end{align}
	where $\lambda$ is determined by the total transmit power constraint $\lVert\bb_{\mathrm{WF}}\rVert_2^2\leq P_\mathrm{tot}$.}
	We then use $\bb_{\mathrm{WF}}$ and $\ba_n$ in \eqref{a vector unequal} for the beamformer update. Similarly, other power distribution strategies, e.g., maxmin power allocation to ensure QoS, can be taken into account with proper setting of $\bb$.
	
	
	
	\subsection{CPR for robust beamforming}\label{sec3-e}
		
	Until now, we have considered the MU-MIMO system with static UEs to clearly explain CPR. In this subsection, we consider the original system with several moving UEs and recover the subscript $i$ that represents the block-fading index to distinguish the outdated and current channels. \red{Then, the channels of $K_\mathrm{s}$ static UEs are fixed over $i$, and the channels of $K_\mathrm{m}$ moving UEs vary over $i$, which makes the BS need to design beamformers without the knowledge of current channel~$\bH_i$.} The outdated channel information at the BS usually reduces the data rate of moving UEs by both the increase of interference and the decrease of desired signal power. For the robustness to the effect of channel uncertainty, we propose CPR with candidate channels (CPR-cc) as another variation of CPR. Without loss of generality, we set the first $K_\mathrm{m}$ UEs are moving among $K=K_\mathrm{m}+K_\mathrm{s}$ UEs.

	CPR-cc uses an extended channel $\overline{\bH}_{i-1}$ that is a concatenation of outdated channel $\bH_{i-1}$ and candidate channel matrices $\bC_{k,i-1}\in\mathbb{C}^{M\times N_\mathrm{c}}$, $k\in\{1,\cdots,K_\mathrm{m}\}$ for moving UEs
	\begin{align}
		\overline{\bH}_{i-1}=[\bH_{i-1}, \bC_{1,i-1},\cdots,\bC_{K_\mathrm{m},i-1}],
	\end{align}
	where $N_\mathrm{c}$ is the number of candidate channels for each moving UE\footnote{It is possible to set different number of candidate channels for each UE depending on the velocity or communication environments, but we use the same number of candidate channels for the sake of simple explanation.}, and the candidate channel matrix $\bC_{k,i-1}$ has columns each of which is a prediction of $\bh_{k,i}$ based on the previous channels $\{\bh_{k,b}\}_{\{b<i\}}$. By containing candidate channels in the extended channel $\overline{\bH}_{i-1}$ and applying CPR, the interference to every candidate channels of moving UEs would be suppressed, and the final beamformer is expected to have less interference on the current channels of moving UEs.
	
	For example, the ordinary CPR beamformer for the $k$-th moving UE $\bff_{k,i}$ is designed to restrain the interference on the subspace spanned by channels of other UEs
	$\mathcal{S}([\bh_{1,i-1},\cdots,\bh_{k-1,i-1},\bh_{k+1,i-1},\cdots,\bh_{K,i-1}])$. When channel varies, however, the current channel subspace $\mathcal{S}([\bh_{1,i},\cdots,\bh_{k-1,i},\bh_{k+1,i},\cdots,\bh_{K,i}])$ may differ from the past channel subspace $\mathcal{S}([\bh_{1,i-1},\cdots, \bh_{k-1,i-1},\bh_{k+1,i-1},$ $\cdots,\bh_{K,i-1}])$
	due to the other $K_\mathrm{m}-1$ moving UEs, and the interference may not be suppressed effectively. On the contrary, $\text{CPR-cc}$ exploits an extended channel matrix and works with the subspace $\mathcal{S}(\overline{\bH}_{-1,i-1})$ where $\overline{\bH}_{-k,i-1}$ represents a matrix that has columns of $\overline{\bH}_{i-1}$ except $\bh_{k,i-1}$ and $\bC_{k,i-1}$. The resulting beamformer of the $k$-th moving UE, then, has lower interference on the other moving UEs.
	
	\red{With the candidate channel matrix $\bC_{k,i-1}$, the current channel of $k$-th moving UE can be represented as
	\begin{align}
		\bh_{k,i} &= \tilde{\bh}_{k,i} + \bee_k,
		\\
		\tilde{\bh}_{k,i} & = \mathrm{proj}\left(\bh_{k,i},\mathcal{S}([\bh_{k,i-1},\bC_{k,i-1}])\right),
	\end{align}
	where $\bee_k\in\mathbb{C}^{M\times1}$ is the effective error. Then, CPR-cc allows to rewrite the interference on the $k$-th moving UE as
	\begin{align}
		\sum_{\substack{\ell=1\\\ell\neq k}}^K \bh_{k,i}^\mathrm{H}\bff_\ell
		=\sum_{\substack{\ell=1\\\ell\neq k}}^K \left(\underbrace{\tilde{\bh}_{k,i}^\mathrm{H}\bff_\ell}_{\substack{\text{interference}\\\text{in control}}} + \underbrace{\bee_k^\mathrm{H}\bff_\ell}_{\substack{\text{interference}\\\text{out of control}}}\right),
		\label{Eq. interference analysis}
	\end{align}
	where $\bff_\ell\in\mathbb{C}^{M\times1}$ is the $\ell$-th UE beamformer. In \eqref{Eq. interference analysis}, the interference in control depends on how much portion of $\bh_{k,i}$ is involved in the subspace $\mathcal{S}([\bh_{k,i-1},\bC_{k,i-1}])$, and this let the extended channel $\overline{\bH}_{i-1}$, which is composed of a proper candidate channel $\bC_{k,i-1}$, be able to largely reduce the interference out of control. Meanwhile, the candidate channel matrix $\bC_{k,i-1}$ that consists of predicted channels of $\bh_{k,i}$ leads each column of $\bC_{k,i-1}$ to be similar to each other.} Therefore, the interference control by exploiting the large subspace $\mathcal{S}(\overline{\bH}_{-k,i-1})$ would cause only marginal beamforming gain loss. The suppressed interference and small decrease of beamforming gain alleviate the data rate degradation of moving UEs.
	
	To be precise, CPR-cc can be conducted similarly with CPR as in Algorithm~\ref{CPR algorithm} by simply substituting Line~7 with
	\begin{align}
		\left(\widehat{\bW}_n\right)_{(:,k)} 
		&= 
		\begin{cases}
			\left( \left(\widehat{\bH}_{n,k}^\mathrm{H}\right)^\dagger  \right)_{(:,1)},\quad 1 \le k \le K_\mathrm{m} ,
			\\
			\left( \left(\widehat{\bH}_{n,k}^\mathrm{H}\right)^\dagger  \right)_{(:,k)},\quad K_\mathrm{m}+1 \le k \le K,
		\end{cases}
	\end{align}
	where
	\begin{align}
		\widehat{\bH}_{n,k}
		&=
		\begin{cases}
			\left[\left([\bh_{k,i-1}, \overline{\bH}_{-k,i-1}]\right)_{(\mathcal{I}_n^{(p)},:)}\right],\   1 \le k \le K_\mathrm{m} ,
			\\
			\left[(\overline{\bH}_{i-1})_{(\mathcal{I}_n^{(p)},:)}\right],\quad  K_\mathrm{m}+1 \le k \le K.
		\end{cases}\label{CPR-cc beamformer update 2}
 	\end{align}
	While any one of candidate channels $(\bC_{k,i-1})_{(:,a)}$, $a\in\{1,\cdots,N_\mathrm{c}\}$ can be used, we adopt the outdated channel $\bh_{k,i-1}$ to construct $\widehat{\bH}_{n,k}$ in \eqref{CPR-cc beamformer update 2} since the BS does not know which candidate channel is more accurate.
	
	With precise channel prediction, the resulting CPR-cc beamformer can improve the beamforming gain. It is important to notice that the accuracy of channel prediction has different meaning when it comes to the interference control in $\text{CPR-cc}$. \red{As it is shown in \eqref{Eq. interference analysis}, the interference control is not about the prediction accuracy of each candidate channel but~about the subspace $\mathcal{S}(\bC_{k,i-1})$ that is spanned by the candidate channels.} In other words, we can use previous two outdated channels $\bh_{k,i-1}$ and $\bh_{k,i-2}$ as candidate channels instead of two or more linear predictions that are based on the previous two channels, where both span the same subspace. The two CPR beamformers, one based on the outdated channels and the other based on the linearly predicted channels, result in the same amount of interference even though one of the linear predictions matches the current channel. Therefore, to derive the full benefit of CPR-cc, a new channel prediction technique is required to produce multiple predicted channels for better interference control, where the purpose of channel prediction becomes totally different from conventional channel prediction techniques.

	
	

	\section{Simulation results}\label{sec4}
	
	In this section, we first analyze \red{the} CPR beamformer and its variations, and then compare some of CPR variations with other beamforming methods in \red{\cite{M.Medra:2018-sum_rate_PAPC,A.Wiesel:2008}, and \cite{M.Ding:2013}. In \cite{M.Medra:2018-sum_rate_PAPC}, a beamformer is designed to maximize a specific weighted sum-rate with a relaxed PAPC $P_{\mathrm{max}}=2P_\mathrm{tot}/M$. For a fair comparison, the resulting beamformer in \cite{M.Medra:2018-sum_rate_PAPC} is additionally scaled down to satisfy the PAPC in \eqref{PAPC}.} The beamforming methods in \cite{A.Wiesel:2008} and \cite{M.Ding:2013} solve multiple optimization problems to design beamformers. The beamforming method in \cite{A.Wiesel:2008} maximizes the sum-rate with zero interference constraints and the PAPC, and the beamforming method in \cite{M.Ding:2013} maximizes \red{the} sum signal-to-interference-plus-noise ratio (SINR) with imperfect channel information and the PAPC. A simply normalized ZF beamformer is also presented as a reference, which is calculated as 
	\begin{align}
		&\bF_\text{nZF}=[\bff_{\text{nZF},1},\cdots,\bff_{\text{nZF},K}],
		\\
		&\bff_{\text{nZF},k} = \nu\frac{\bff_{\text{ZF},k}}{\left\lVert \bff_{\text{ZF},k} \right\rVert_2}, \quad k\in\{1,\cdots,K\},
		\\
		&[\bff_{\text{ZF},1},\cdots,\bff_{\text{ZF},K}]
		= \left(\bH_{i-1}^\mathrm{H}\right)^\dagger,
	\end{align}
	where $\nu$ is the normalization factor to satisfy the PAPC. We set $\nu$ to have the maximum magnitude as
	\begin{align}
		\nu &= \min_{m\in\{1,\cdots,M\}} \sqrt{P_\mathrm{ant}} \left( \sum_{k=1}^{K} \left\lvert \frac{(\bff_{\text{ZF},k})_{(m)}}{\left\lVert \bff_{\mathrm{ZF},k}\right\rVert_2 }\right\rvert^2 \right)^{-\frac{1}{2}}.
	\end{align}

	We consider $M$ transmit antennas at the BS and $K_\mathrm{m}$ moving UEs and $K_\mathrm{s}$ static UEs. The channel instantiation interval is set as $t=40$ ms, and the velocity of moving UEs are assumed to be the same as $v$. The Doppler frequency then becomes $f_{D,k}=vf_c/c$ with carrier frequency $f_c=2.3$ GHz and the speed of light $c=3\cdot 10^8$ m/s. For the channel model, we set the large-scale fading term $\beta_k=z_k/x_k$ where $x_k\sim\mathrm{U}[1,5]$ models the pathloss effect and $z_k$ models the shadowing effect following a log-normal distribution with standard deviation $8$~dB \cite{H.Q.Ngo:2013}.
	Each $\angle r_k$ for channel correlation matrix is uniformly distributed in $[0,2\pi)$ with $|r_k|=0.6$ as in \cite{H.Kim:2019}. 
	The magnitude of transmit symbol is set to be $|s_{k,i}|=1$, and the BS signal-to-noise ratio (SNR) is $\frac{P_\mathrm{tot}}{\sigma^2}$.
	\red{FC-CPR is designed with $p_\mathrm{init}=0.699$, $\Delta_p=0.05$, and $p_\mathrm{max}=0.999$, and CPR is designed with $p=p_\mathrm{init}$ or $p=p_\mathrm{max}$ where each value is optimized numerically.} For CPR-cc, two outdated channels $\bH_{i-1}$ and $\bH_{i-2}$ are considered as $N_\mathrm{c}=2$ candidate channels of moving UEs, and the sum-rate is obtained~as
	\begin{align}
		\sum_{k=1}^K \log_2\left(1+ \frac{|\bh_{k,i}^\mathrm{H}\bff_{k}|^2}  {\sigma^2+\left\lvert \sum_{\ell\neq k} \bh_{k,i}^\mathrm{H}\bff_{\ell} \right\rvert^2} \right),
	\end{align}
	\red{where the beamformers $\bff_k$, $k\in\{1,\cdots,K\}$ are designed before estimating the current channel~$\bH_{i}$.}

	\begin{figure}[!t]
		\centering
		\includegraphics[width=.95\linewidth]{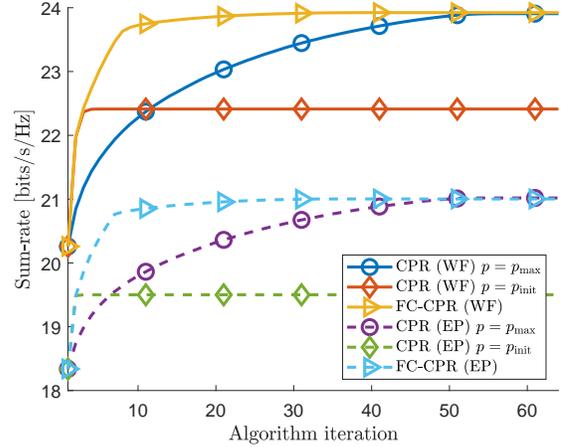}
		\caption{Sum-rate over algorithm iteration with ${M=64}$, ${K_\mathrm{m}=0}$, $K_\mathrm{s}=8$, SNR $=0$ dB.}
		\label{data rate over algorihtm iteration}
	\end{figure}
	\begin{figure}[t]
		\centering
		\subfloat[$K_\mathrm{m}=0$, $K_\mathrm{s}=4$]{	\includegraphics[width=.95\linewidth]{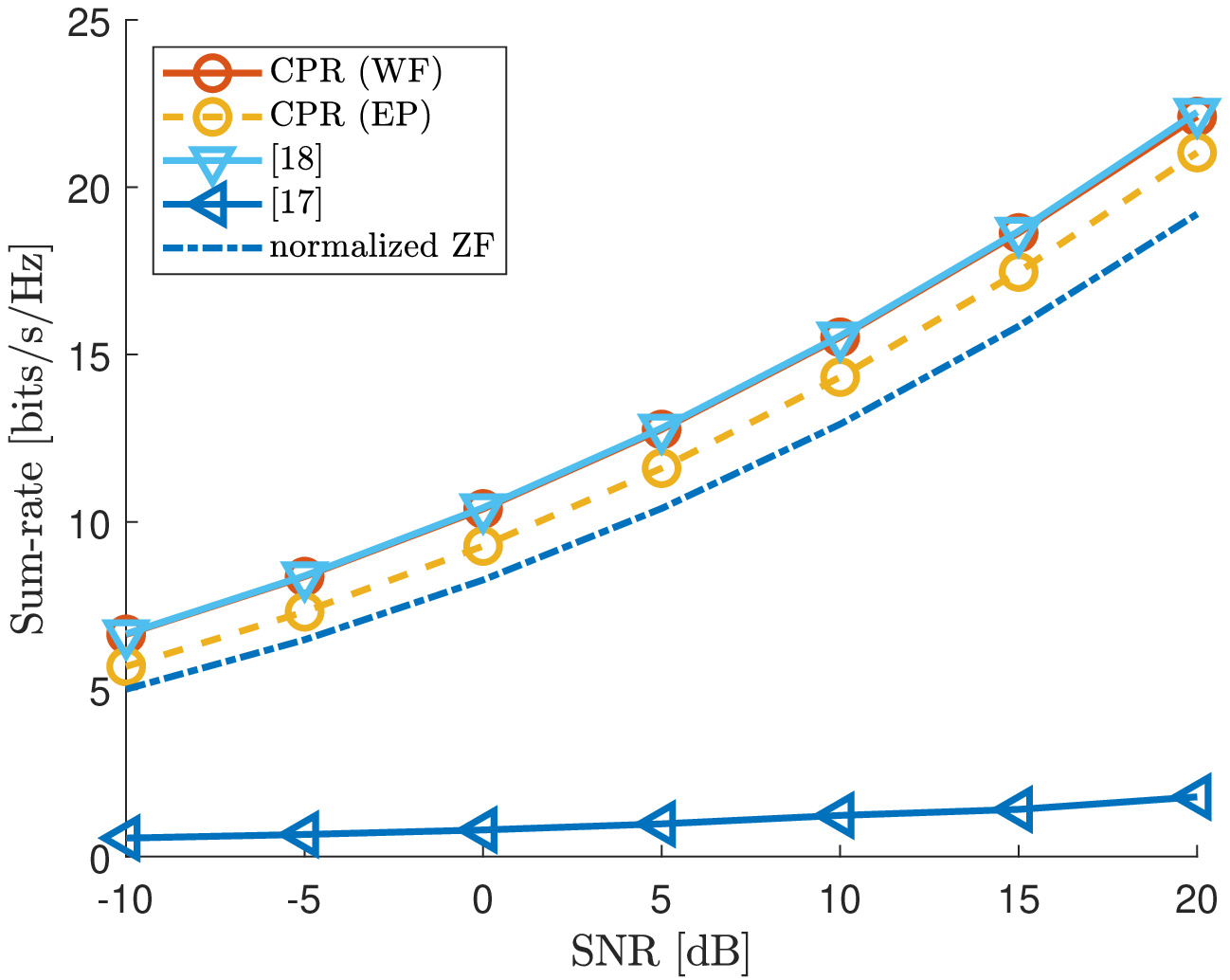}
			\label{sum-rate over SNR 1}
		}\\
		\subfloat[$K_\mathrm{m}=0$, $K_\mathrm{s}=8$]{
			\includegraphics[width=.95\linewidth]{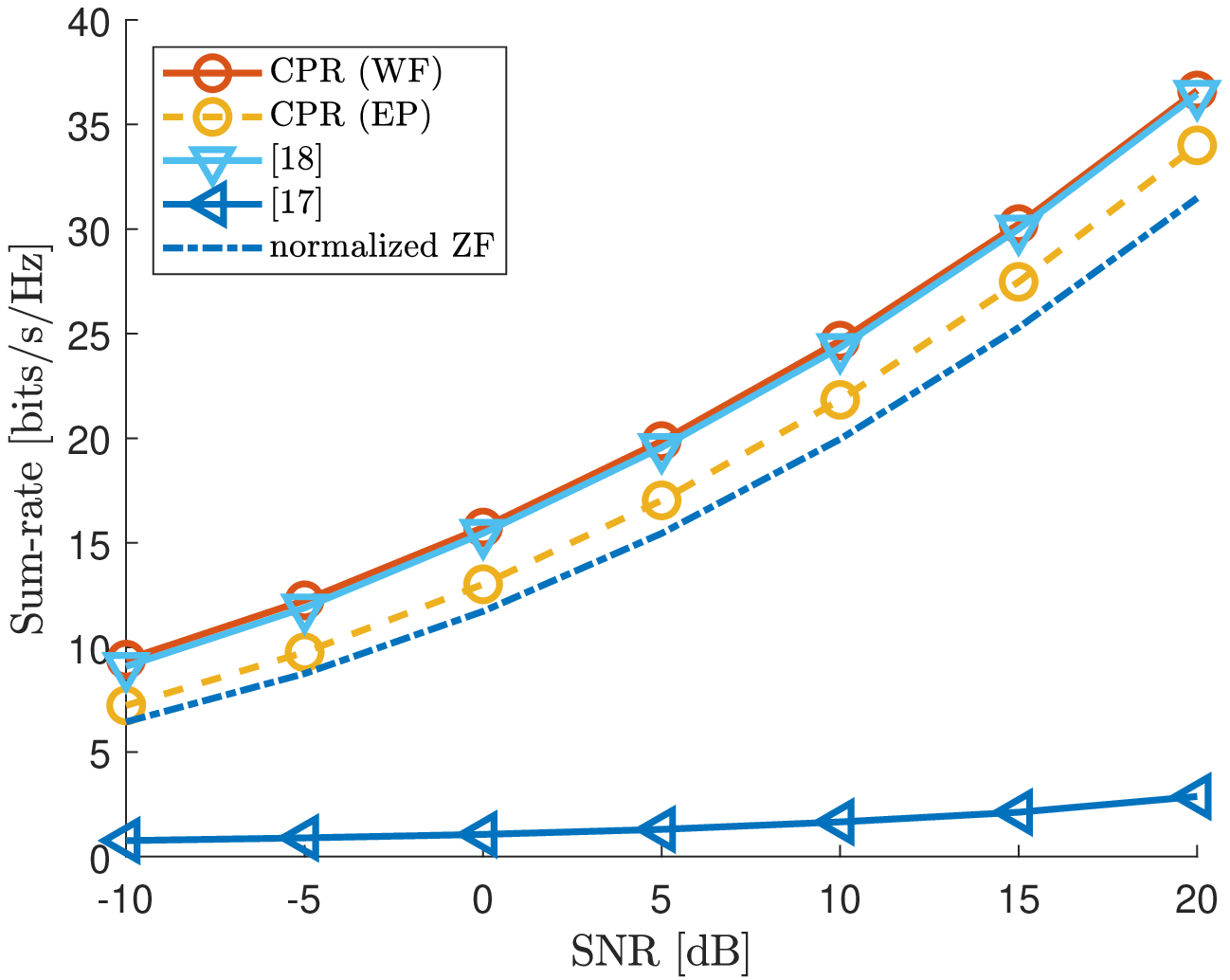}
			\label{SubFig. sum-rate stationary K40 M32}
		}
		\caption{Sum-rate over SNR with ${M=16}$.}
		\label{Fig. sum-rate stationary}
	\end{figure}
	
	\begin{figure}[!t]
		\centering
		\subfloat[$M=16$]{
			\includegraphics[width=.95\linewidth]{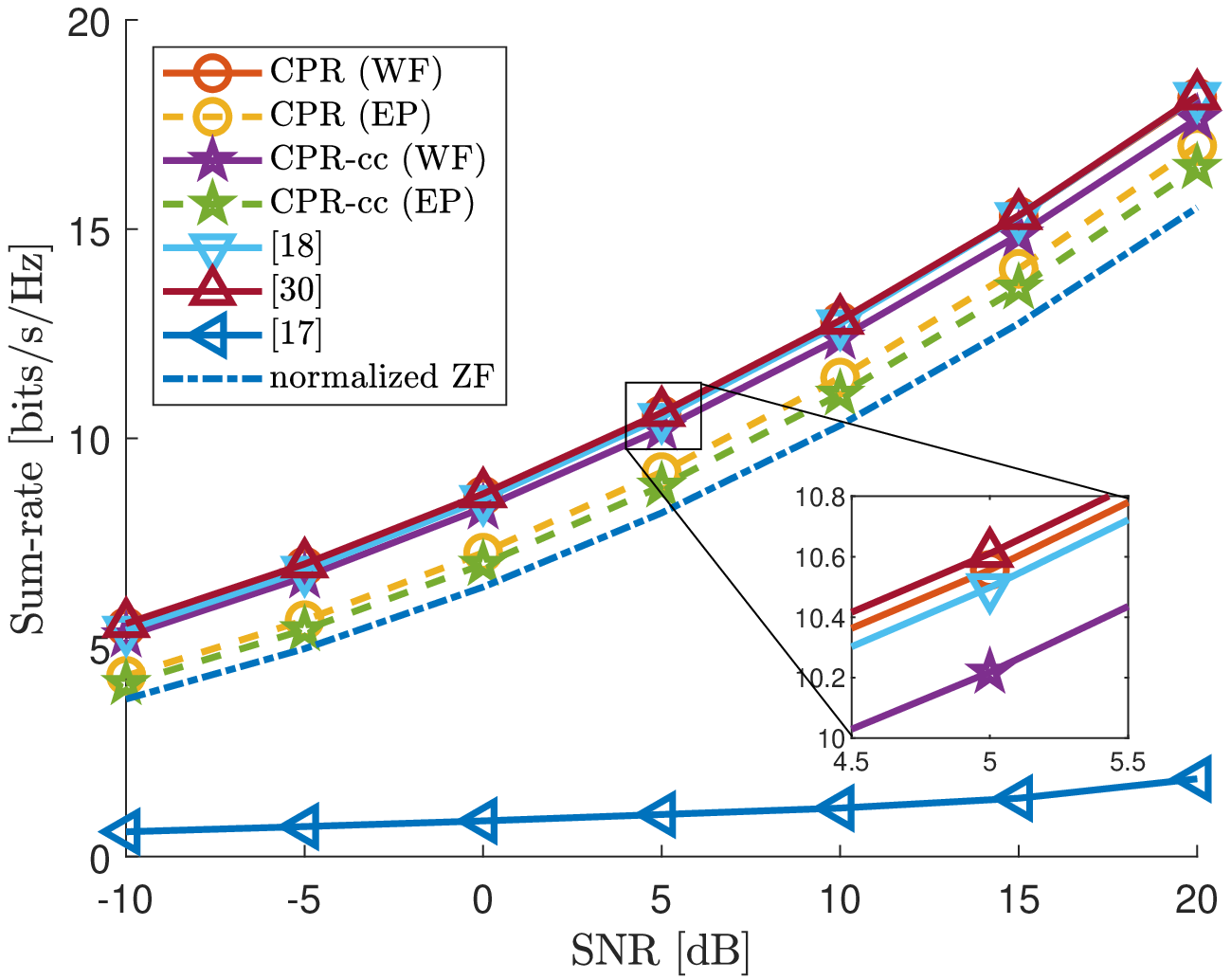}
			\label{sum-rate over SNR K31 M16}
		}\\
		\subfloat[$M=32$]{
		\includegraphics[width=.95\linewidth]{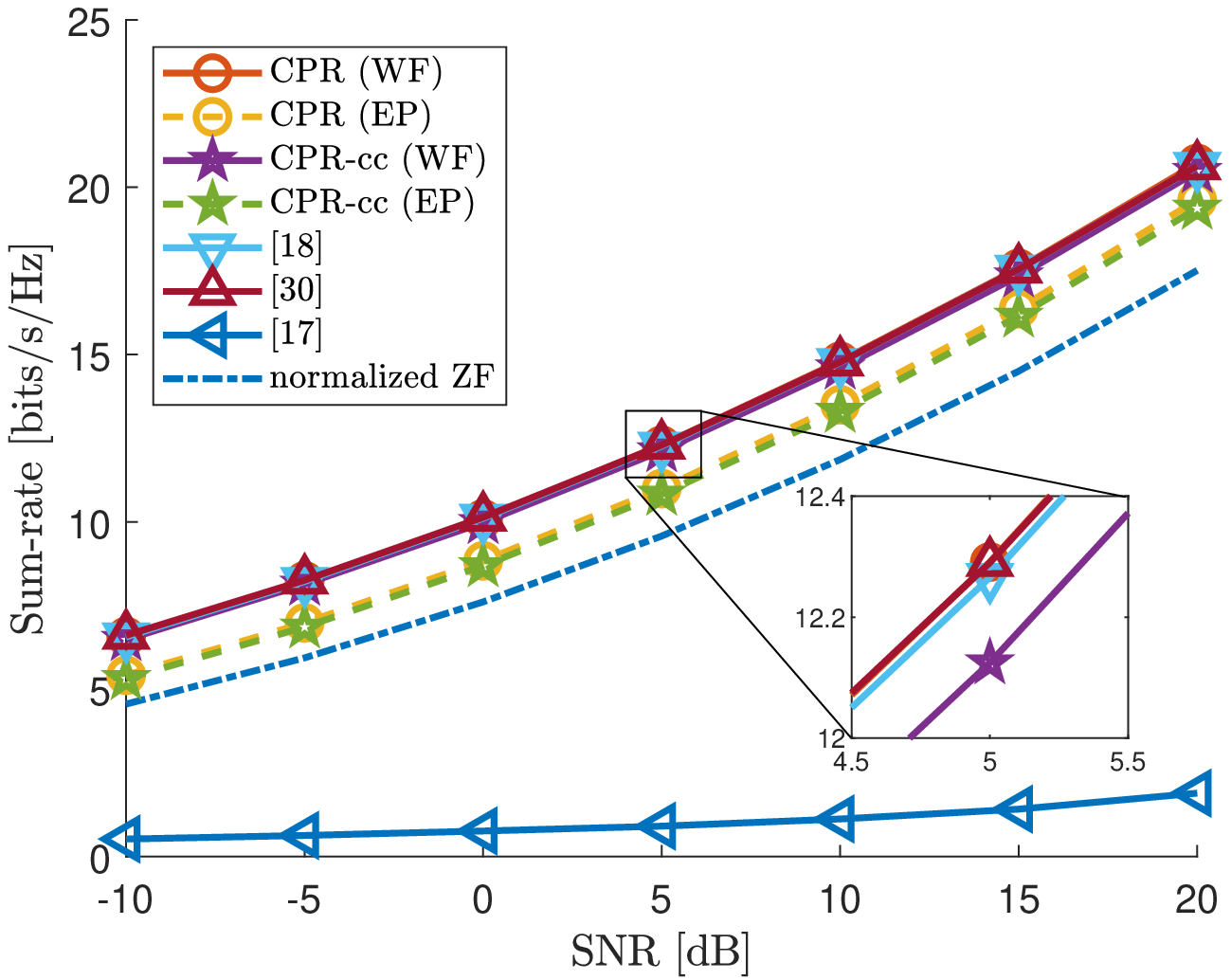}
		\label{SubFig. sum-rate K71 M16}
		}
		\caption{Data rate over SNR with $K_\mathrm{m}=1$, $K_\mathrm{s}=3$, $v=3\ \mathrm{km/h}$.}
		\label{data rate over SNR}
	\end{figure}
	\begin{figure}[t!]
		\centering
		\includegraphics[width=.95\linewidth]{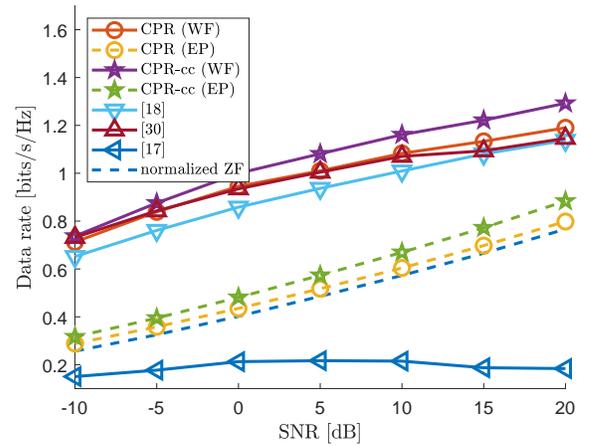}
		\caption{Data rate of moving UE over SNR with $M=16$, $K_\mathrm{m}=1$, $K_\mathrm{s}=3$, $v=3\ \mathrm{km/h}$.}
		\label{moving UE data rate over SNR K31 M8}
	\end{figure}

	\subsection{Algorithm convergence}\label{sec4-a}
	
	CPR iteratively updates a beamformer, and \red{$p$ balances the maximum performance and the convergence speed. To improve both the convergence speed and the maximum performance, we proposed FC-CPR, and numerical results are depicted in Fig. \ref{data rate over algorihtm iteration} to assess the sum-rate over algorithm iterations.} Depending on the power distribution strategy, (EP) and (WF) are denoted to represent the equal power distribution and the water-filling power distribution, respectively. Regardless of distribution strategy, \red{CPR with small $p=p_\mathrm{init}$ quickly converges but provides a low sum-rate. On the contrary, CPR with large $p=p_\mathrm{max}$ slowly converges to a high sum-rate with around 50 iterations. FC-CPR achieves both the high sum-rate and fast convergence by approaching the high sum-rate of CPR with large $p=p_\mathrm{max}$ near the 8-th iteration. By the virtue of its simple structure, CPR can be operated with large $p=p_\mathrm{max}$ under a moderate complexity.
	Hence, in the next subsection, only the results of CPR with large $p=p_\mathrm{max}$ are depicted for the readability of figures.}

	\red{Another factor that differs the maximum sum-rate is the power distribution strategy. The water-filling distribution gives a higher sum-rate than equal power distribution, and it is natural since the water-filling is designed to raise the sum-rate.} When the minimum data rate has \red{a} high priority, however, the equal power distribution would be better than the water-filling distribution. The two distribution strategies are shown as examples, and a proper strategy can be set depending on a specific purpose.
		
	\subsection{Data rate comparison}\label{sec4-b}
	
	
%
	\begin{figure}[t]
		\centering
		\includegraphics[width=.95\linewidth]{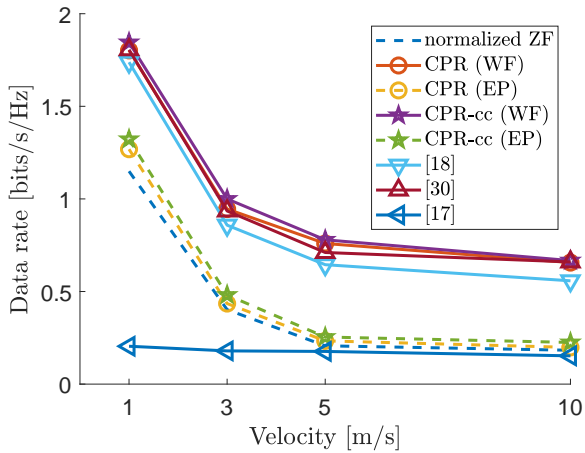}
		\caption{Data rate of moving UE over velocity with ${M=16}$, ${K_\mathrm{m}=1}$, ${K_\mathrm{s}=3}$, SNR $=10$ dB.}
		\label{moving UE data rate over v}
	\end{figure}

	We consider two scenarios for data rate comparison, one with all static UEs and the other with a moving UE. The first scenario is considered in Fig. \ref{Fig. sum-rate stationary} without the results of CPR-cc and the beamformer in \cite{M.Ding:2013}, which are designed to compensate the channel uncertainty. The sum-rate of CPR (WF) and the beamformer in \cite{A.Wiesel:2008} is similarly high and shows \red{a} remarkable gain compared to the normalized ZF beamformer. This clearly shows that CPR (WF) can achieve the same data rate with \cite{A.Wiesel:2008}, which is based on complicated optimization problems, with significantly less complexity.
	The sum-rate of CPR (EP) are between the simply normalized ZF beamformer and the above beamformers. \red{The beamformer in \cite{M.Medra:2018-sum_rate_PAPC} is designed to maximize a specific weighted sum-rate, which weights the data rate of each UE with the inverse of squared norm of channel vector, and the sum-rate with equal weights for UEs are lower than simply normalized ZF beamformer due to the relaxed PAPC.}
	
	The simulation results of the second scenario are in Fig.~\ref{data rate over SNR} with the velocity of moving UE $3$ km/h. \red{The beamformers in Fig.~\ref{Fig. sum-rate stationary} still have the same trend in Fig.~\ref{data rate over SNR} while the sum-rate of beamformer in \cite{M.Ding:2013} is close to that of CPR~(WF). The sum-rate of CPR-cc~(WF) is little lower than that of CPR~(WF) since CPR-cc is to improve the performance of moving UE, not the sum-rate. The simple use of outdated channel degraded the sum-rate of CPR-cc (WF), and the use of better channel prediction method, which is out of scope of this paper, would be able to raise the sum-rate of CPR-cc (WF) as well.
	The robustness is assessed by the data rate of moving UE in Fig.~\ref{moving UE data rate over SNR K31 M8}.} 
	The beamformer in \cite{M.Ding:2013}, which is a robust version of the beamformer in \cite{A.Wiesel:2008}, provides higher data rate than the beamformer in \cite{A.Wiesel:2008}. As SNR increases, however, the data rate of the beamformer in \cite{M.Ding:2013} falls below that of CPR~(WF), i.e., even CPR~(WF) is more robust to the channel uncertainty than the beamformer in \cite{M.Ding:2013}. \red{Although the sum-rate of CPR-cc~(WF) is little lower than that of CPR~(WF), CPR-cc~(WF) gives the highest data rate for the moving UE, providing robustness to the channel uncertainty.}
	
	In Fig. \ref{moving UE data rate over v}, the data rate of moving UE is assessed according to the velocity $v$. Over the velocity, the data rate of moving UE decreases, while CPR-cc (WF) still gives the highest data rate, and CPR~(WF) outperforms the beamformer in \cite{M.Ding:2013} for all range of velocity of interest. 

	

	\section{Conclusion}\label{sec5}
	
	We proposed a beamformer design to effectively exploit transmit power under the PAPC. CPR develops a beamformer by iteratively adding up extra beamformers. All the processes are conducted by linear operations without solving intricate optimization problems, which ensures low computational complexity. In addition, the simple structure of CPR makes it highly flexible and allows CPR to be easily adapted to various scenarios. As examples, we designed several variations of CPR in the perspective of convergence speed, sum-rate maximization, and robustness for the channel uncertainty. The simulation results verified that CPR and its variations satisfy their design purposes.
	
	\section*{Appendix I: Proof of Lemma 1}
	
	\begingroup
	\allowdisplaybreaks
	\red{
		First, the beamforming gain of the $k$-th UE is 
		\begin{align}
			&\left\lvert \bh_k^\mathrm{H} (\bF_n)_{(:,k)} \right\rvert^2 
			\notag\\
			&= \left\lvert \bh_k^\mathrm{H} (\bF_{n-1}+\bW_n\bA_n)_{(:,k)} \right\rvert^2
			\notag\\
			&=\left\lvert \bh_k^\mathrm{H} (\bF_{n-1})_{(:,k)} \right\rvert^2 
			\notag\\
			&\qquad 
			+ 2\cdot \mathrm{Real} \left\{\left(\bh_k^\mathrm{H} (\bF_{n-1})_{(:,k)}\right)^* (\ba)_{(k)}\bh_k^\mathrm{H}(\bW_n)_{(:,k)} \right\}  
			\notag\\
			&\qquad 
			+ \left\lvert(\ba)_{(k)}\bh_k^\mathrm{H}(\bW_n)_{(:,k)} \right\rvert^2
			\notag\\
			&\stackrel{(a)}{>} \left\lvert \bh_k^\mathrm{H} (\bF_{n-1})_{(:,k)} \right\rvert^2 ,
		\end{align}
		where $(a)$ is by the coefficients $(\ba_n)_{(k)}$ that are designed in \eqref{a vector} to align the products of channel and two beamformers $\angle\left(\bh_k^\mathrm{H}(\bF_{n-1})_{(:,k)}\right)=\angle\left((\ba_n)_{(k)}\bh_k^\mathrm{H}(\bW_{n})_{(:,k)}\right)$. With the alignment, $\mathrm{Real} \left\{\left(\bh_k^\mathrm{H} (\bF_{n-1})_{(:,k)}\right)^* (\ba)_{(k)}\bh_k^\mathrm{H}(\bW_n)_{(:,k)} \right\}$ equals to $\left(\bh_k^\mathrm{H} (\bF_{n-1})_{(:,k)}\right)^*(\ba)_{(k)} \bh_k^\mathrm{H}(\bW_n)_{(:,k)}\ge0$. In addition, $\widehat{\bW}_n$ is the ZF beamformer and provides a positive beamforming gain.
	}
	\endgroup
	
	\red{
		Next, the interference of the $\ell$-th beamformer $({\bF_n})_{(:,\ell)}$ to the $k$-th UE channel $\bh_k$ is
		\begin{align} 
			\bh_k^\mathrm{H} ({\bF_n})_{(:,\ell)}
			&
			=\underbrace{\bh_k^\mathrm{H} ({\bF_{n-1}})_{(:,\ell)}}_{\text{existing interference}} + \underbrace{(\ba)_{(\ell)} \bh_k^\mathrm{H}({\bW_n})_{(:,\ell)}}_{\text{additional interference}},
		\end{align}
		where $k\in\{1,\cdots,K\}$, $\ell\in\{1,\cdots,K\}$, and $k\neq \ell$. The additional interference disappears as
		\begin{align}
			(\ba)_{(\ell)} \bh_k^\mathrm{H}({\bW_n})_{(:,\ell)}
			&=(\ba)_{(\ell)} \sum_{m=1}^M(\bh_k)_{(m)}^* ({\bW_n})_{(m,\ell)}
			\notag\\
			&
			=(\ba)_{(\ell)} \sum_{m\in\mathcal{I}_n^{(p)}} (\bh_k)_{(m)}^* ({\bW_n})_{(m,\ell)} 
			\notag\\
			&\qquad
			+ (\ba)_{(\ell)} \sum_{m\notin\mathcal{I}_n^{(p)}} (\bh_k)_{(m)}^* ({\bW_n})_{(m,\ell)}
			\notag\\
			&
			=(\ba)_{(\ell)} (\bh_k)_{(\mathcal{I}_n^{(p)})}^\mathrm{H} (\widehat{\bW}_n)_{(:,\ell)}
			\notag\\
			&\stackrel{(a)}{=} 0,
		\end{align}
		where $(a)$ is by the fact that the beamformer $\widehat{\bW}_n$ is the ZF beamformers of $(\bH)_{(\mathcal{I}_n^{(p)},:)}$. 
	}
	%

	\bibliographystyle{IEEEtran}
	\bibliography{CPR_references}

\end{document}